\documentclass[11pt,twoside]{article}
\usepackage{asp2004}
\usepackage{epsf}

\begin{document}

\title{Eclipse mapping of the stream and disk flickering components
	in V2051~Oph}
\author{R. Baptista and A. Bortoletto}
\affil{Departamento de F\'{\i}sica - CFM, UFSC, 88040-900, 
	Florian\'opolis, Brazil}

\begin{abstract}
We report on the investigation of the spatial distribution of the 
flickering sources in the dwarf nova V2051~Oph with eclipse mapping 
techniques.  Low-frequency flickering originates in the gas stream and 
is related to the mass transfer process, whereas high-frequency 
flickering arises in the accretion disk and is probably connected 
to magneto-hydrodynamic turbulence.
\end{abstract}

V2051~Oph is distinguished among other eclipsing dwarf nova by its 
remarkable flickering activity (amplitude $\ga 30$ per cent).  
The large flickering amplitude and short orbital period make this binary 
an ideal target for flickering studies with eclipse mapping techniques.

Time-series of CCD photometry of V2051 Oph in the B-band were obtained 
at the Laborat\'orio Nacional de Astrof\'{\i}sica (Brazil), from 1998 to 
2002, while the star was in quiescence.
The data comprise 36 eclipse light curves obtained with the same 
instrument and telescope, which ensures a high degree of uniformity 
to the data set.
The data can be grouped in two different brightness levels, named the 
'faint' and 'bright' states.  The differences in brightness are caused 
by long-term variations in the mass transfer rate from the secondary star.
The white dwarf is hardly affected by the long-term changes. Its
flux increases by only 10 per cent from the faint to the bright state,
whereas the disk flux raises by a factor of 2.

We applied the `single' and `ensemble' methods \citep[see][]{bruch2000}
to the set of light curves of V2051~Oph to derive the orbital dependency 
of its steady-light, long-term brightness changes, low- and 
high-frequency flickering components.
The `ensemble' method samples flickering at all frequencies. But because 
the power spectrum density of the flickering is well described by a 
power-law $P(f) \propto f^{-\alpha}$, an ensemble curve 
is dominated by the low-frequency flickering components. 
On the other hand, the `single' method produces curves which sample
the high-frequency flickering.
Thus, the combination of both methods opens the possibility not only 
to probe the location of the flickering sources but also to separate
low- and high-frequency components of the flickering.

Eclipse maps of the steady-light show asymmetric brightness 
distributions with enhanced emission along the ballistic stream
trajectory, in a clear evidence of gas stream overflow.
The comparison between the steady-light maps of the faint and bright
states suggests that the quiescent disk of V2051~Oph responds to changes 
in mass transfer rate in an homologous way.

Our flickering mapping analysis reveal the existence of two different
sources of flickering in V2051~Oph, which lead to variability at
distinct frequencies.  The low-frequency flickering arises mainly in 
the overflowing gas stream and is connected to the mass transfer 
process.  It maximum emission occurs at the position of closest approach 
of the gas stream to the white dwarf, and its spatial distribution 
changes in response to variations in mass transfer rate.
In the bright state there is additional contribution from an uneclipsed
component of 8 per cent of the total flux, which may indicate the 
development of a vertically-extended clumpy or turbulent wind from the 
inner disk regions.
Unsteady mass transfer or turbulence generated after the shock between
the stream and the disk material may be responsible for this 
stream flickering component.

The high-frequency flickering is produced in the accretion disk.
It is spread over the disk surface with a radial distribution similar 
to that of the steady-light maps and it shows no evidence of emission 
from the hot spot, gas stream or white dwarf.  This disk flickering 
component has a relative amplitude of about 3 per cent of the 
steady-light, independent of disk radius and brightness state. 
If the disk flickering is caused by fluctuations in the energy 
dissipation rate induced by MHD turbulence \citep{ga}, its relative 
amplitude lead to a viscosity parameter $\alpha_{cool}= 0.1-0.2$ at all 
radii for the quiescent disk of V2051~Oph \citep{bb04}.  
This is comparable to the inferred viscosity in outburst $\alpha_{hot}
\simeq 0.14$ \citep{santos} -- in contradiction with a fundamental 
prediction of the disk instability model (namely, $\alpha_{hot} \ga 5 
\; \alpha_{cool}$).

Previous studies suggested that flickering may originate from two
separate sources, the hot spot and a turbulent inner disk region 
\citep{w95,bruch2000}. 
Our flickering mapping experiment reveals a more complex situation,
in which flickering arising
from the inner disk regions may in fact have the same origin as 
the hot spot flickering, namely, the mass transfer process.
Stream flickering from the inner disk regions may be largely 
dominant over the disk flickering if gas stream overflow occurs 
in a quiescent dwarf novae, because the contribution from the 
disk flickering in this case is small (since the disk itself is 
relatively faint).  Also, although the hot spot contributes to 
the flickering, it is a minor source in comparison to the
inner gas stream in this situation.  This seems to be the case
in V2051~Oph.
On the other hand, if the disk flickering amplitude is a fixed 
fraction of the steady disk light, it might become the dominant
flickering source and may appear as arising mainly from the innermost 
disk regions in a high mass transfer, nova-like system as the surface 
brightness of the bright, opaque disks of these systems 
decreases sharply with radius.

Finally, we note that the measurement of the disk flickering component 
yields a novel and independent way to estimate the disk viscosity 
parameter $\alpha$.  The large value found in this first experiment 
seems to favor the mass-transfer instability model \citep[see][]{w95}.

\end{document}